\documentstyle[aps,preprint,eqsecnum]{revtex}
\begin{document}
  \draft
\title{On the Interplay of Monopoles and Chiral Symmetry Breaking in
Non-Compact Lattice QED}
\author{John B. Kogut}
\address{Department of Physics,
University of Illinois at Urbana-Champaign\\
1110 West Green Street,
Urbana, IL 61801-3080}
\author{K. C. Wang}
\address{School of Physics,
University of New South Wales\\
P.O. Box 1,
Kensington, NSW 2203, Australia}
\maketitle

\begin{abstract}
Non-compact lattice QED is simulated for various numbers of fermion
species $N_f$ ranging from 8 through 40 by the exact Hybrid Monte Carlo
algorithm.  Over this range of $N_f$, chiral symmetry breaking is found
to be strongly correlated with the effective monopoles in the
theory.  For $N_f$ between 8 and 16 the chiral symmetry breaking and monopole
percolation transitions are second order and coincident.  Assuming powerlaw
critical behavior, the correlation length exponent for the chiral transition is
identical to that of monopole percolation.  This result supports the
conjecture that monopole percolation ``drives" the nontrivial chiral
transition.
  For $N_f$ between 20 and 32, the monopoles experience a first order
condensation transition coincident with a first order chiral transition.  For
$N_f$ as large as 40 both transitions are
strongly suppressed.  The data at large $N_f (N_f \mathrel {\mathpalette \vereq
>} 20)$ is interpreted in terms of a strongly interacting monopole gas-liquid
transition.
\end{abstract}
\pacs{     }
\vfill
\pagebreak

\section{Introduction}

The interplay of monopole and fermion dynamics has been studied both
analytically and computationally in quantum field theory for some time.  In the
context of Grand Unified model building, the existence of monopole solutions of
the field equations and the subsequent interactions of the monopole with
the theory's fermions has led to several interesting phenomena including the
existence of exotic fermion condensates and the Callan-Rubakov effect
\cite{Callan}.  The
Dirac condition plays a crucial role in these discussions and in some cases it
guarantees that qualitatively new, non-perturbative effects occur.

It is, therefore, of some interest when monopoles play a role in a broader
context.  In particular, various models studied in lattice gauge theory afford
new glimpses into monopole physics, since the \underline{second quantized}
field theory of monopoles becomes accessible.  A classic example is the
confinement-deconfinement transition in abelian lattice gauge theory which is
now understood to be driven by monopole condensation \cite{Banks}.  It came
as some
surprise, however, when effective monopoles were discovered in non-compact
lattice QED \cite{Hands}, and the percolation transition of these objects
was seen to be the
same as four dimensional bond percolation \cite{Kocic}.  The lattice itself
allows such
objects to have finite action, but one would expect naively that they would
decouple in the theory's continuum limit.  But this depends on the scaling
properties of the monopole percolation transition.  When non-compact lattice
QED
is coupled to fermions in the traditional, explicitly gauge invariant fashion
of Schwinger (through U(1) phases) one finds that the monopole percolation and
the chiral symmetry transitions are coincident \cite{Kogut}.  Even more
tantalizing is the
fact that the correlation length exponents for both transitions also coincide,
so monopoles could survive the continuum limit of the chiral transition and the
chiral transition itself may represent an interacting continuum field theory
in four dimensions.  Needless to say, it will prove extraordinarily hard to
establish such a scenario purely numerically \cite{Gockeler}.
\vfill
\pagebreak
It is the purpose of this paper to elucidate the interplay of monopole and
fermion dynamics in non-compact lattice QED by studying the system's phase
transitions as $N_f$, the number of fermion species, is varied.  Since the
fermion-monopole interaction strength is determined by the Dirac quantization
condition, $N_f$ is the only natural variable available.  We are not in a
position to
make a quantitative study of each theory's continuum limit.  All our
results will
be of a semi-quantitative or even qualitative sort.  We think that they are
interesting and have content, nonetheless.  Our lattice sizes will be modest
(typically
$10^4$) as will our bare fermion masses (typically of order 0.05 and
greater in lattice units).  It
turned out that finite size effects are particularly large when $N_f$ is
large  so smaller bare fermion masses cannot be sensibly studied on $10^4$
lattices.  This fact will also force our chiral symmetry breaking studies to be
done rather far from the chiral limit, and therefore they are not as
quantitative as one might hope.

Since this paper continues a large body of work discussed in detail elsewhere,
it will rely on definitions and formulas presented before
\cite{Hands,Kocic,Kogut}.  The reader should
consult those references for background.  Quantities such as the chiral
condensate $\langle{\overline\psi}\psi\rangle$, the monopole susceptibility
$\chi$ and the monopole percolation order parameter $M$ should be familiar and
will not be reviewed here.  Equation of State, scaling laws and critical
indices should also be familiar to the reader. Our notation will be the
same as past publications.

We begin with an overview of our results.  For $N_f$ equal 2 and 4 we will
rely on
past, more quantitative studies \cite{Kogut}.  The $N_f = 8\ {\rm and}\
16$ simulations were
done on $10^4$ lattices, with selected simulations on $12^4, 14^4\ {\rm and}\
16^4$ lattices to check for finite size effects.  Accurate studies of the
chiral
condensate showed that only bare fermion masses greater or equal to 0.05 (in
lattice units) are free of finite size effects.  This result implies that some
past studies of lattice QED at large $N_f$ were not under quantitative
control \cite{Dagotto}.
Data taken at bare fermion masses of .05, .06, .07, .08, .09 and .10, and at
couplings ranging from $\beta = .21$ to .14 in steps of $\Delta\beta =
.005$, are
consistent with the hypothesis that there is a chiral transition at
$\beta_c = .17(1)$ with powerlaw critical singularities.  The critical indices
for the chiral transition are consistent with those measured more precisely for
the $N_f = 2$ and 4 theories previously.  Measurements of the monopole
percolation observables also indicate a second order phase transition  at
essentially the same coupling, $\beta_m = .180(5)$, with critical indices
characteristic of conventional four dimensional bond (or site) percolation.
The
coincidence of the chiral and monopole percolation transitions has been noted
before in the $N_f = 2\ {\rm and}\ 4$ theories \cite{Kogut}.  Even more
intriguing than this
is the fact that both transitions may share the same correlation length index
$\nu$.  Precise measurements of monopole percolation in four dimensions have
strongly suggested the exact result $\nu = {2 \over 3}$ \cite{Kocic}.
Measurements of the
chiral equation of state presented here for $N_f = 8$ and elsewhere for $N_f =
2\  {\rm and}\  4$ give the critical indices $\delta = 2.2(1)\ {\rm and}\
\gamma = 1.0(1)$.  If we assume that the critical point has powerlaw
singularities with conventional properties, then the critical indices should
satisfy the hyperscaling relations, and $\delta\  {\rm and}\ \gamma$ determine
all of them.  The hyperscaling relations read,
\begin{eqnarray}
2-\alpha &=& d\nu \nonumber \\
2\beta_{mag}\delta - \gamma &=& d\nu \nonumber\\
\beta_{mag} &=& {\nu \over 2} (d - 2 + \eta)  \\
2\beta_{mag} + \gamma &=& d\nu \nonumber
\end{eqnarray}
Then the values $\delta = 2.2\ {\rm and}\  \gamma = 1$
imply $\beta_{mag} = .83, \nu =
.67, \alpha = -.67, \eta = .50\ {\rm and\ }
\Delta \equiv \beta_{mag}\delta = 1.83$.
The intriguing result of this exercise is that measurements of the chiral
exponents and the hypothesis of hyperscaling predict that the correlation
length exponents $\nu$ of both transitions are identical.  This implies that
the
monopoles are relevant degrees of freedom at the chiral transition and since
they scale identically, the monopoles should survive in the continuum limit of
the chiral transition.  It might be accurate to say that monopole percolation
``drives" the chiral transition and the chiral transition defines an
interacting, continuum field theory because it  ``inherits'' the non-mean-field
correlation length critical index $\nu = {2 \over 3}$ from monopole
percolation.  We shall see in the text through analysis of our measurements of
the order parameter $\langle {\overline \psi}\psi\rangle$ that this
interpretation fits the computer simulation data very well.  However, other
hypotheses, such as the possibility that the chiral transition is described by
a
logarithmically trivial Nambu Jona-Lasinio model, might fit the data
adequately as well.  It would require considerably more computer power to
separate the monopole picture of the transition from other possibilities just
on the basis of numerical fits.  One reason for this difficulty is the fact
that finite size effects grow large as $N_f$ increases and small bare fermion
masses close to the chiral limit cannot be simulated on lattices of practical
proportions like $10^4$ or even $16^4$.  For this reason the emphasis in this
paper will be different, although elsewhere the $N_f = 2$ and 4 models are
being
simulated on even larger lattices with even better statistics.

We consider the $N_f = 12, 16, 20, 24, 32\ {\rm and}\ 40$ models here, and
measure $\langle {\overline \psi}\psi\rangle$ and monopole observables to see
if the correlation between these observables persist at all $N_f$.  We shall
see
that while the character of the transitions change qualitatively as $N_f$
increases, the two transitions remain strongly correlated.  To avoid finite
size effects we were forced to simulate a relatively large bare fermion mass $m
= 0.10$ on $10^4$ lattices.  Therefore, many of our conclusions are just
qualitative.  Luckily, qualitative changes were seen in the data as $N_f$
varied so the study remained useful.  We found that as $N_f$ was increased from
8 to 24, the chiral and monopole percolation transitions both shifted to
stronger critical couplings but they remained coincident and
apparently second order.  However, at $N_f \approx 24$ and $m = 0.10$ both
transitions displayed jumps suggesting first order behavior.  (The reader
should be careful not to overlook our caveats expressed above in these
remarks--simulations at smaller $m$ and larger lattices are really necessary to
make such statements.)  At $N_f = 24$, the chiral condensate $\langle
{\overline\psi}\psi\rangle$ and the monopole percolation order parameter $M$
display ``discontinuous" jumps between couplings $\beta = 0.08\ {\rm and}\
.085$.  Increasing $N_f$ even further to 32 simply enhances the sizes of the
apparent discontinuities.  The fact that the character of the transition for
$N_f$ between 8 and 24 is different from that for $N_f$ between 24 and 40 is
supported by measurements of the monopole concentration (density).  For small
$N_f$ where the monopole transition is percolation, the concentration of
monopoles is expected to be small and smooth through the transition.  The
simulations show this clearly.  However, our simulations at
$N_f = 24$ and especially $N_f = 32$ show that the concentration jumps
``discontinuously" at the chiral transition, strongly suggesting a first order
transition between a dilute ``gaseous state" of monopoles and a fairly
dense ``liquid" state.  For example, at $N_f = 32$ the monopole concentration
jumps from 0.10 (approx.) at $\beta = .055\ {\rm to}\ 0.34\ {\rm at}\ \beta =
0.050$.  Since the transition shows up in the monopole concentration it is
accurate to call it a ``condensation" transition.   Since c = 0.34 is a
substantial density, it is very tantalizing to view the transition as a first
order gas-liquid transition.  Apparently increasing $N_f$ effects the
monopole core energy and/or the
monopole-monopole interactions and thereby induces a gas-liquid transition.
It would be
interesting to complement the computer results with some analytic calculations.

Of course this physical picture of the transition needs further
substantiation.  One element of it that we could test here was the expectation
that if fermion-induced forces were effecting the monopole dynamics and leading
to a gas-liquid transition, then if $N_f$ were taken truly large free monopoles
would never appear in the system.  In fact, we confirmed this point at $N_f =
40$.  The simulation showed that the monopole concentration and the monopole
percolation susceptibility and the monopole percolation order parameter
all remained strongly suppressed and flat as the coupling varied.  The chiral
condensate was similarly suppressed.  However, the plaquette showed strong
dependence on the coupling $\beta$ suggesting that a transition remains in the
model with a divergent specific heat, but it is unrelated to the monopole or
chiral properties of the model.  The small values of M and c indicate that the
monopoles remain bound in tight pairs for all coupling $\beta$.  Under these
circumstances one would not expect them to induce chiral symmetry breaking and
our simulations are consistent with that fact.

The phase diagram ($N_f\ vs.\ \beta$) that we are advocating here agrees
qualitatively with that of Azcoiti and collaborators \cite{Azcoiti}.  Their
work emphasizes the
theory's specific heat, while ours emphasizes monopole and chiral dynamics.  We
believe that they are two views of the same physics and the qualitative
features
we are interested in and can deal with fairly reliably are identical.

The body of this paper is organized as follows.  In Sec. 2 the $N_f = 8$ theory
is discussed in detail.  In Sec. 3 we turn to the $N_f = 12, 16, 20\  {\rm
and}\
24$ data, and show that the $N_f = 24$ data displays monopole condensation.
In Sec. 4 we turn to the $N_f = 32\ {\rm and}\ 40$ data which shows that for
truly large $N_f$ the monopole and chiral activities in the theory are strongly
suppressed.

\section{The $N_{\lowercase{f}} = 8$ Simulation}

We used our Hybrid Monte Carlo code for non-compact QED to explore the eight
flavor, $N_f = 8$, model just as we studied the $N_f = 4$ case more
quantitatively in an earlier publication.  The reader should consult our
extensive $N_f = 2\ {\rm and}\ 4$ studies for details of the algorithm and
the definitions of various chiral and monopole observables
\cite{Hands,Kocic,Kogut}.  Since this paper is
looking for qualitative trends and is a contribution in a long series we will
not repeat formulas, definitions and past observations.  Rather, our emphasis
will be on results, plots and an emerging physical picture.

To gain some understanding of the chiral transition at $m = 0.0$ we measured
$\langle{\overline \psi}\psi\rangle$ for bare fermion masses ranging from $.03\
{\rm to}\  .10$ and couplings $\beta = {1/e^2}$ ranging from $.20\ {\rm
to}\ .14$.  The data is shown in Table 1.  Several hundred trajectories of the
Hybrid Monte Carlo code were required to achieve the statistical accuracy
indicated in the table.  Since the lattice size is relatively small, $10^4$, we
must be careful about finite size effects in the data, especially at small
values of m.  Therefore, we did limited simulations on $12^4, 14^4\ {\rm and}\
16^4$ lattices.  The data is shown in Tables 2, 3 and 4.  Comparing Tables 1
and 2 we see evidence on the weak coupling side of the chiral transition,
$\beta = .20 - .18$, for numerically significant finite size effects for the
lowest fermion mass, $m = .03$.  Over this range of parameters,
$\langle{\overline \psi}\psi\rangle$ is relatively suppressed on the smaller
lattice, which is the expected finite size/finite temperature effect.  However,
comparing Tables 1-4 we see that the finite size effects are within our
statistical error bars at $m = .05$, except perhaps at the weakest coupling
$\beta = .200$.  Therefore, in the analysis that follows only the $10^4$ data
for m ranging from .05 to .10 will be used.

We will assume that the chiral transition is well-described by a second order
phase transition with powerlaw singularities.  Other hypotheses could be tried
here and some would probably be fairly successful since our data covers only
relatively large m values and because every fitting hypothesis is accompanied
by several free parameters.  We will pursue the powerlaw hypothesis here
because it is simple and because it is definitely appropriate for monopole
percolation \cite{Kocic}.  Given this, the data should satisfy the equation
of state (EOS),
\begin{equation}
\langle{\overline \psi}\psi\rangle /m^{1\over\delta} = f (\Delta\beta
/\langle{\overline \psi}\psi\rangle^{1 \over \beta_{mag}})
\end{equation}
where $\delta\ {\rm and}\ \beta_{mag}$ are the usual critical indices,
$\Delta\beta = \beta_c - \beta$, and this form of the EOS has been used
extensively elsewhere.  Eq. (2.1) simplifies at the critical point and reduces
to the scaling law of the order parameter $\langle{\overline\psi}\psi\rangle$
as
the symmetry breaking field m is turned on,
\begin{equation}
\langle{\overline\psi}\psi\rangle = Am^{1 \over\delta}\ \ \ \ \ , \beta =
\beta_c
\end{equation}
We found that Eq.(2.2) is a particularly effective way to determine $\delta$
and $\beta_c$ which then can be used in the EOS to find the universal scaling
function $f$ and the critical index $\beta_{mag}$ away from the transition.  In
Fig. 1 we plot $-1/ln\langle{\overline\psi}\psi\rangle vs. -1/ln(m)$ for the
data of Table 1 $(m \geq .05)$.  At $\beta = \beta_c$ these lines should be
straight with the slope ${1 \over \delta}$ and should pass through the origin.
We see that this hypothesis works well at $\beta_c = .17$ in Fig. 1 for $m =
.05 - .10$.  The lower masses are subject to finite size effects, as discussed
above, and must be discarded.  The slope of the $\beta_c = .17$ line in Fig. 1
gives $\delta = 2.2(1)$.  Powerlaw fits of the $\beta = .17$ data to Eq. (2.2)
are excellent indicating that the numerical evidence for the powerlaw
hypothesis is perfectly consistent with the numerical data.

We note that the result $\delta = 2.2(1)$ is consistent with the results found
at $N_f = 2\ {\rm and}\ 4$, assuming powerlaw singularities at the critical
point.  Those data also gave the susceptibility index $\gamma = 1.0(1)$ which
by the hyperscaling relation $\beta_{mag} = \gamma/(\delta -1)$ predicted
$\beta_{mag} = .83(7)$, the magnetic critical index.  This motivated us to try
the EOS Eq. (2.1) with $\beta_{mag} = .83$.  The result is shown in Fig. 2.
The data fall on a universal scaling function $f$ rather well, although the
quality of the data and the resulting universal curve are not comparable to our
$N_f = 2\ {\rm and}\ 4$ results which came from larger lattices and smaller
values of $m$.  However, if we compare the EOS for $N_f = 8$ in Fig. 2 to the
analogous figures in the
$N_f = 2\ {\rm and}\ 4$ publications, we see that even the universal
function $f$
as well as the critical indices $\delta\ {\rm and}\ \beta_{mag}$ are consistent
with their independence of $N_f$.  One interpretation of this result is that
monopole percolation drives each transition, as discussed in the Introduction
above, and fermion feedback does not effect the percolation critical behavior
as long as $N_f$ is not too large.  More evidence for this scenario will be
presented below when monopole observables are presented and
analyzed.  There is no doubt, however, that other more mundane explanations of
these systematics could be presented.  For example, it could be that all the
$N_f \neq 0$ theories are logarithmically trivial and have the scaling
properties of Nambu Jona-Lasinio (NJL) models.  If this hypothesis is true, the
reason for the deviation of $\delta\ {\rm and}\ \beta_{mag}$ from their
mean-field values of 3 and 1/2 is the presence of scale-breaking logarithms
in the NJL equation of state.  The limited accuracy of our $N_f = 8$ makes it
pointless to pursue alternative fits here--given a few new parameters as would
occur in NJL fits, this could certainly be done.  Rather we shall investigate
just qualitative features of the models with higher $N_f$ and accumulate
additional evidence for strong correlations between the chiral and monopole
activities in each model.  This will then provide ``supporting, circumstantial
evidence" for the monopole-driven-chiral transition physical picture we are
developing.

Next we accumulated monopole percolation data for the $N_f = 8$ theory on a
$10^4$ lattice at various $m$.  The data for $\chi$, the monopole
susceptibility, and $M$, the monopole percolation order parameter, are
presented
in Table 5.  We see that the peak of the susceptibility $\chi$ occurs at a
coupling between $.175\ {\rm and}\ .180$ for $m = .03$, and it moves to
slightly
weaker coupling, $.19$, as $m$ increases to $.10$.  Our estimate of $\beta_c =
.17$ for the chiral transition refers, of course, to the $m = 0$ chiral
limit.  So, within uncertainties due to finite size effects, the chiral and
monopole percolation transitions are coincident, as we found with better
numerical control for $N_f = 2\ {\rm and}\ 4$.  It is important to determine if
the peaks in $\chi$ on the $10^4$ lattice are indicative of a real transition.
To obtain some evidence for this result and to measure some critical indices,
we repeated the measurements summarized in Table 5 on $10^4, 12^4, 14^4\ {\rm
and}\ 16^4$ lattices at $m = .05$.  The data is given in Table 6.  We see that
the peaks grow with $L$.  According to finite size scaling, the peak heights
should grow as,

\begin{equation}
\chi_{max} (L) \sim L^{\gamma_{mon}/\nu_{mon}}
\end{equation}
where $\gamma_{mon}\ {\rm and}\ \nu_{mon}$ are the susceptibility and
correlation length exponents for the monopole transition.  In addition, the
order parameter at the coupling $\beta_{max}$ where $\chi$ peaks for each $L$
should scale to zero as,
\begin{equation}
M(\beta_{max}(L)) \sim L^{-\beta_{mon}/\nu_{mon}}
\end{equation}
We test these scaling predictions in
Fig. 3 and find that the data supports powerlaw scaling with the indices,
\begin{eqnarray} \gamma_{mon}/\nu_{mon} = 2.25(3) \nonumber\\
\beta_{mon}/\nu_{mon} = .875 (80)
\end{eqnarray}
These are exactly the critical
indices of ordinary four dimensional percolation.  Four dimensional percolation
indices satisfy hyperscaling relations and Eq. (2.5) then predicts $\nu_{mon} =
{2\over 3}$.  This is the correlation length scaling index discussed in the
Introduction.  Its coincidence with the correlation length exponent of the
chiral
transition is crucial to the monopole-percolation-driven-chiral-transition
physical picture.

In summary, our $10^4$ numerical results are compatible with the idea that the
$N_f = 8$ chiral transition is physically indistinguishable from the $N_f = 2\
{\rm and}\ 4$ chiral transitions.  If we assume powerlaw critical
singularities, then the physical picture of monopole percolation driving the
chiral transition is also defensible because the couplings of the transitions
coincide as do their correlation length indices.

\section{Monopole Condensation at $N_{\lowercase{f}} = 24$}
We next increased $N_f$ in our Hybrid Monte Carlo code and simulated the $N_f =
12, 16, 20\ {\rm and}\ 24$ models on $10^4$ lattices with $m=0.10$.  A
relatively large bare fermion mass was chosen to control finite size effects.
The relatively large symmetry-breaking field will smooth out the chiral
transition and make quantitative investigations impossible.  However,
qualitative changes in the dynamics of the model will be seen.  The reader
should understand, however, that we cannot predict the precise $N_f$ values
where qualitative changes occur.  More simulations at smaller bare fermion
masses on larger lattices will be needed for that.

The simulation data for the average plaquette $P$, the chiral condensate
$\langle{\overline\psi}\psi\rangle$ and the monopole percolation order
parameter $M$ are shown in Fig. 4 for $N_f = 12\ {\rm and}\ 16$.  The
transition region between small $\langle{\overline\psi}\psi\rangle$ (or $M$),
and large $\langle{\overline\psi}\psi\rangle$ (or $M$) shifts toward stronger
coupling and the transition sharpens, somewhat.  The shift toward stronger
coupling is a consequence of dynamical fermion screening and has been seen in
many contexts before.  In Fig. 5, we show the same plots for $N_f = 20\ {\rm
and}\ 24$.  Now there are suggestions that for each $N_f$ the order parameters
$\langle{\overline\psi}\psi\rangle$ and $M$ jump at the same coupling from
small to larger values.  This is particularly persuasive for $N_f = 24$ where
we see signs of discontinuities at $\beta = .105(5)$.  Perhaps this
qualitative effect is more visual in Fig. 6 where the $N_f = 24$ and $N_f = 8$
data for $m = 0.10$ are plotted and we have added the monopole concentration
(density) ``$c$" to the list of observables.  The chiral condensate, monopole
concentration and average plaquette each appear to jump for $\beta =
.0825(25)$.  Certainly for strong coupling, $\beta < .0825$, their slopes are
much greater than their slopes at weak coupling, $\beta > .0825$.  By contrast,
the same set of observables are smooth in the plot of the
$N_f = 8$ data.  Of course, there is a transition in the $N_f = 8$ data,
but it does not show up clearly at relatively large values of $m$, except in
the
monopole percolation observables $M$ and $\chi$.  In fact, we plot the monopole
percolation susceptibilities $\chi$ for the $N_f = 8\ {\rm and}\ 24$ theories
at
$m = 0.10$ in Fig. 7.  Strong peaks are seen for both $N_f$ values with the
width of the $N_f = 24$ peak considerably reduced, again indicating the
relative sharpness of the $N_f = 24$ transitions.

Perhaps the clearest indication that the dynamics of the $N_f = 24$ model is
qualitatively different from the $N_f = 8$ case comes from the monopole
concentration.  As seen in Tables 7 and 8, of $N_f = 8$ and $N_f = 24$ data at
$m = 0.10$ on $10^4$ lattices, the monopole concentration ``jumps" in the $N_f
=24$ case while it is perfectly smooth through the percolation transition in
the $N_f = 8$ case.  This hints at the fact that the monopoles are condensing
in the $N_f = 24$ theory and the ground state for $\beta < .0825(25)$ is a
monopole condensate, perhaps resembling the strong coupling, confining vacuum
of the compact $U(1)$ lattice QED model.  Since the monopole concentration is
small for $\beta > .0825(25)$ and jumps to a distinctly larger value for $\beta
< .0825(25)$, we may be seeing signs of a first-order gas-liquid monopole
condensation transition.  The monopole activation energy is proportional to ${1
\over e^2} = \beta$ and it is relatively small here compared to the small $N_f$
models.  As the coupling is increased through .0825(25) a first order monopole
condensation transition into a monopole liquid is triggered where a relatively
dense monopole ensemble is produced.  It would be
interesting to study the monopole dynamics through correlation functions in
this
condensed state and compare them to similar simulations in pure compact QED.

\section{Monopole and Chiral Suppression at $N_{\lowercase{f}} = 40$}
In this survey of $N_f$, we next turned to the $N_f = 32$ model.  The data is
presented in Table 9 (for $m = 0.10$ and $10^4$ lattices, as usual) and it is
plotted in Fig. 8.  Jumps are seen in all observables for $N_f = 32$ at a
coupling $\beta = .05125(125)$.  On the strong coupling side of the transition,
$M, \langle{\overline\psi}\psi\rangle$ and $c$ are saturated.  The average
plaquette has also jumped at $\beta = .05125(125)$, and is growing rapidly in
the strong coupling phase.  A first-order monopole condensation transiton is
very apparent.

We finally increased $N_f$ to 40 in order to see the effects of extreme fermion
screening.  Table 10 and Fig. 8 resulted--the monopole and chiral observables
are almost completely suppressed!  Throughout the entire range of couplings
$\langle{\overline\psi}\psi\rangle$ remains near its weak coupling value.  Both
of the percolation observables, $\chi\ {\rm and}\ M$, are strongly
suppressed and
are slightly smaller at $\beta = .01$ than at $\beta = .02$.  The average
plaquette $P$ rapidly increases over this range of $\beta$, however, probably
indicative of a persistent specific heat anomaly, as discussed more
quantitatively by Azcoiti and  \cite{Azcoiti}.  Our interest in this
result is again the strong correlation between the monopole and chiral
observables.  The fact that they are both deeply suppressed, even while the
average plaquette indicates considerable ``disorder" in the ground state, is
supportive of the physical picture which contends that the effective monopoles
are essential in the model's chiral dynamics at all $N_f$.

\section{Concluding Remarks}

In this survey of $N_f$ we have found that chiral and monopole dynamics are
strongly correlated in every case.

\subsection{Small $N_f$}

The monopole transition is a second order percolation transition without
condensation.  If the chiral transition is assumed to be characterized by
 powerlaw singularities, satisfying hyperscaling, then it was coincident
with monopole percolation and the correlation length indices of the two
transitions were identical.

\subsection{Intermediate $N_{\lowercase{f}}$}

The monopole transition becomes a first order condensation phenomenon.  The
chiral transition is coincident and also first order.

\subsection{Large $N_f$}

The monopole and chiral observables are strongly suppressed, and there are no
transitions in these quantities.  The average plaquette, is rapidly varying as
a function of coupling nonetheless.

In summary, it may be worthwhile to pursue some aspects of the dynamics found
here in more detail.  The nature of the chiral transition for small $N_f$ is a
primary goal, since it may define an interacting field theory which is strongly
coupled at short distances.  The nature of the  field theory and the
role of effective monopoles in it would be interesting to understand.  The
monopole condensate at intermediate $N_f$ and its ``liquid" properties would be
interesting to clarify through correlation functions.

\section{Table Captions}

\begin{itemize}
\begin{enumerate}

\item Chiral Condensate data for $N_f = 8$ theory on $10^4$
lattice at various couplings $\beta$ and bare fermion masses $m$.

\item Same as Table 1 except on a $12^4$ lattice.

\item Same as Table 1 except on a $14^4$ lattice.

\item Same as Table 1 except on a $16^4$ lattice.

\item Same as Table 1 except for monopole susceptibility $\chi$ and order
parameter $M$.

\item  Monopole percolation measurements in $N_f = 8$ theory at $m = .05$ on
$L^4$ lattice for $L = 10, 12, 14\  {\rm and}\  16$.

\item  $N_f = 8$ data for $m = 0.10$ on a $10^4$ lattice.

\item  Same as Table 7 except $N_f = 24$.

\item  Same as Table 7 except $N_f = 32$.

\item  Same as Table 7 except $N_f = 40$.
\end{enumerate}
\end{itemize}

\begin{table}
\caption{Chiral Condensate, $N_f = 8$, L = 10}
\label{1.}
\begin{tabular}{ddddddddd}
$\beta/m$    & .03 & .04 & .05 & .06 & .07 & .08 & .09 & .10\\
\tableline
.200 & .1066(5) & .1396(7) & .1677(7) & .1947(6) & .2148(7) & .2368(8) &
.2545(7) & .2739(7)\\
.195 & .1139(7) & .1468(9) & .1778(8) & .2056(7) & .2270(8) & .249(1) & .269(1)
& .2835(7)\\
.19 & .1246(6) & .160(1) & .1903(8) & .2160(7) & .2388(8) & .260(1) & .280(1) &
.2958(7)\\
.185 & .1370(7) & .173(1) & .2036(8) & .2290(8) & .252(1) & .274(1) & .293(1) &
.3098(8)\\
.18 & .147(1) & .188(1) & .2179(8) & .247(1) & .268(1) & .288(1) &
.305(1) &.3232(8)\\
.175 &  .167(1) & .209(1) & .236(1) & .261(1) & .286(1) &
.304(1) & .319(1) & .3364(9)\\
.17 & .181(1) & .219(1) & .256(1) & .279(1) & .302(1) & .321(1) & .335(1) &
.352(1)\\
.165 & .203(1) & .242(1) & .274(1) & .300(1) & .322(1) & .335(1) & .353(1) &
.369(1)\\
.16 & .232(1) & .265(1) & .297(1) & .320(1) & .340(1) & .358(1) & .371(1) &
.382(1)\\
.155 & .262(1) & .293(1) & .323(1) & .341(1) & .360(1) &.377(1) & .386(1) &
.399(1)\\
.15 & .291(2) & .319(1) & .344(1) & .364(1) & .380(1) & .394(1) &
.407(1) & .417(1)\\
.145 & .324(2) & .349(1) & .371(1) & .386(1) & .405(1) & .416(1) & .425(1) &
.434(1)\\
.14 & .355(2) & .377(1) & .398(1) & .412(1) & .424(1) & .435(1) & .444(1) &
.452(1)
\end{tabular}
\end{table}

\begin{table}
\caption{Chiral Condensate, $N_f = 8, L = 12$}
\label{2.}
\begin{tabular}{ddd}
$\beta/m$ & .03 & .05\\
\tableline
.200 & .1123(5) & .1723(8)\\
.190 & .1280(8) & .1916(8)\\
.185 & .1388(7) & .2030(6)\\
.180 & .1516(7) & .2185(6)\\
.175 & .1657(7) & .2344(8)\\
.170 & .1824(8) & .2520(7)\\
.160 & .229(1)  & .2949(9)\\
.150 & .289(1)  & .345(1)\\
.140 & .355(1)  & .396(1)
\end{tabular}
\end{table}

\begin{table}
\caption{Chiral Condensate, $N_f = 8, L =14$}
\label{3.}
\begin{tabular}{dd}
$\beta/m$ & .05\\
\tableline
.185 & .2032(4)\\
.180 & .2188(4)\\
.170 & .2356(5)
\end{tabular}
\end{table}

\begin{table}
\caption{Chiral Condensate, $N_f = 8, L = 16$}
\label{4.}
\begin{tabular}{dd}
$\beta/m$ & .05\\
\tableline
.185 & .2043(3)\\
.180 & .2184(3)\\
.170 & .2340(4)
\end{tabular}
\end{table}

\begin{table}
\squeezetable
\caption{Monopole Data, $N_f = 8, L = 10$}
\label{5.}
\begin{tabular}{ccccccccccc}
& \multicolumn{2}{c}{$m = 0.03$}& \multicolumn{2}{c}{$m =
0.04$}&\multicolumn{2}{c}{$m = .05$} & \multicolumn{2}{c}{$m = .06$} &
\multicolumn{2}{c}{$m = .10$}\\
\tableline
$\beta$ & $\chi$ & $M$ & $\chi$ & $M$ & $\chi$ & $M$ & $\chi$ & $M$ & $\chi$ &
$M$ \\
\tableline
.20 & 19.75(16) & .041(1) & 21.98(22) & .045(1) & 23.45(32) & .051(1) &
25.15(27) & .053(1) & 34.69(56) & .088(2)\\
.195 & 24.1(2) & .050(1) & 26.48(30) & .055(1) & 28.55(30) & .060(1) &
30.89(46) & .078(2) & 45.4(9) & .125(3)\\
.19 & 30.9(4) & .071(1) & 34.00(50) & .076(2) & 36.44(52) & .084(2) & 40.35(61)
& .099(3) & 51.6(1.5) & .221(6)\\
.185 & 39.7(7) & .106(3) & 41.58(74) & .123(3) & 48.4(.97) & .137(4) &
48.11(1.11) & .180(5) & 37.7(1.97) & .383(6)\\
.180 & 48.5(9) & .152(4) & 55.53(1.4) & .180(4) & 49.6(1.8) & .266(6) &
45.6(2.2) & .330(6) & 19.94(1.6) & .535(5)\\
.175 & 46.0(1.8) & .298(6) & 42.3(1.9) & .351(6) & 31.11(1.99) & .434(6) &
27.18(1.7) & .473(6) & 8.31(98) & .652(3)\\
.17 & 28.1(2.3) & .457(6) & 18.6(1.4) & .519(5) & 11.12(66) & .586(3) &
9.04(74)
& .623(3) & 3.70(10) & .747(2)\\
.165 & 13.8(1.1) & .579(4) & 7.67(52) & .649(3) & 5.45(17) & .692(2) & 4.08(12)
& .727(2) & 2.11(3) & .810(1)\\
.16 & 5.19(17) & .709(2) & 3.99(11) & .736(2) & 2.94(6) & .772(2) & 2.28(3) &
.798(1) & 1.33(2) & .859(1)\\
.155 & 2.48(3) & .789(1) & 2.09(3) & .809(1) & 1.58(2) & .837(1) & 1.32(1) &
.855(1) & .078(1) & .899(1)\\
.15 & 1.46(2) & .848(1) & 1.191(2) & .867(1) & 0.95(1) & .883(1) &
0.81(1) & .895(1) & 0.50(1) & .9270(6)\\
.145 & .082(1) & .897(1) & 0.69(1) & .908(1) & 0.55(1) & .920(1) & &  & .032(1)
& .9483(4)\\
.14 & 0.50(1) & .927(1) & 0.42(1) & .937(1) & 0.32(1) & .946(1) & & & 0.21(1) &
.9639(3)
\end{tabular}
\end{table}

\begin{table}
\caption{Monopole Observable Scaling, $N_f = 8$}
\label{6.}
\begin{tabular}{ccccccccc}
$\beta/L$ & \multicolumn{2}{c}{10} & \multicolumn{2}{c}{12} &
\multicolumn{2}{c}{14}  & \multicolumn{2}{c}{16}\\
\tableline
& $\chi$ & $M$ & $\chi$ & $M$ & $\chi$ & $M$ & $\chi$ & $M$ \\
\tableline
.20 & & & 25.8(3) & .0267(2) \\
.19 & & & 48.0(6) & .058(1)\\
.185 & & & 66(1) & .105(3) & 83(1) & .076(2) & 104(2) & .061(1)\\
.18 & 50(2) & .266(6) & 75(3) & .228(5) & 107(3) & .197(4) & 142(4) & .180(3)\\
.175 & & & 42(3) & .409(4) & 40(3) & .415(4) & 40(2) & .412(2)\\
.17 & & & 11.9(4) & .575(2)\\
.16 & & & 2.89(4) & .773(1)\\
.15 & & & .937(8) & .886(1)\\
.14 & & & .359(4) & .945(1)
\end{tabular}
\end{table}

\begin{table}
\caption{$N_f = 8$ Data}
\label{7.}
\begin{tabular}{cccccc}
$\beta$ & P & $\langle{\overline\psi}\psi\rangle$ & $\chi$ & $M$ & $c$\\
\tableline
.21 & .961.(2) & .2550(8) & 21.9(2) & .046(1) & .1104(2)\\
.205 & .972(2) & .2643(8) & 27.0(3) & .060(1) & .1093(2)\\
.20 & .983(2) & .2739(7) & 34.7(6) & .088(2) & .1163(2)\\
.195 & .993(2) & .2835(7) & 45.4(9) & .125(3) & .1232(2)\\
.19 & 1.005(2) & .2958(7) & 52(2) & .221(6) & .1309(2)\\
.185 & 1.023(2) & .3098(8) & 38(2) & .383(6) & .1396(3)\\
.18 & 1.040(2) & .3232(8) & 20(2) & .535(5) & .1482(2)\\\
.175 & 1.110(2) & .3364(9) & 8.3(9) & .652(3) & .1573(2)\\
.17 & 1.151(2) & .352(1) & 3.7(1) & .747(2) & .1682(2)\\
.165 & 1.180(2) & .369(1) & 2.11(3) & .810(1) & .1791(2)\\
.16 & 1.196(2) & .382(1) & 1.33(2) & .859(1) & .1900(2)\\
.155 & 1.271(2) & .399(1) & .78(1) & .899(1) & .2033(2)\\
.15 & 1.310(2) & .417(1) & .50(1) & .9270(6) & .2163(2)\\
.145 & 1.372(2) & .434(1) & .32(1) & .9483(4) & .2299(2)\\
.14 & 1.464(2) & .452(1) & .21(1) & .9639(3) & .2451(2)
\end{tabular}
\end{table}

\begin{table}
\caption{$N_f = 24$ Data}
\label{8.}
\begin{tabular}{dcdddd}
$\beta$ & $P$ & $\langle{\overline\psi}\psi\rangle$ & $\chi$ & $M$ & $c$\\
\tableline
.12 & .799(1) & .183(1) & 11.2(1) & .019(1) & .0730(2)\\
.11 & .845(1) & .193(1) & 14.8(2) & .029(2) & .0847(2)\\
.10 & .915(1) & .211(1) & 25.1(3) & .053(2) & .1002(2)\\
.095 & .955(1) & .222(1) & 34.2(7) & .088(3) & .1089(2)\\
.09 & 1.019(1) & .238(1) & 50(2) & .197(7) & .1211(3)\\
.085 & 1.097(1) & .258(1) &  26(3) & .476(7) & .1356(3)\\
.08 & 1.336(2) & .329(1) & 1.36(3) & .853(2) & .1794(5)\\
.07 & 2.77(1) & .516(2) & .0076(5) & .9983(2) & .3354(8)
\end{tabular}
\end{table}

\begin{table}
\caption{$N_f = 32$ Data}
\label{9.}
\begin{tabular}{dcdccd}
$\beta$ & $P$ & $\langle{\overline\psi}\psi\rangle$ & $\chi$ & $M$ & $c$\\
\tableline
.09 & .7366(6) & .1524(4) & 8.29(4) & .0024(6) & .052(1)\\
.08 & .8100(7) & .1586(4) & 9.68(5) & .0102(9) & .061(1)\\
.07 & .934(1) & .1687(5) & 11.90(8) & .0218(9) & .073(1)\\
.06 & 1.276(2) & .1867(6) & 20.2(3) & .045(1) & .092(1)\\
.0575 & 1.509(2) & .1954(9) & 29.2(9) & .061(4) & .100(1)\\
.055 & 1.692(1) & .2070(7) & 41.7(9) & .146(6) & .110(1)\\
.0525 & 2.151(1) & .2339(7) & 17.4(9) & .509(6) & .130(1)\\
.05 & 4.639(4) & .509(1) & .0048(2) & .9989(1) & .335(1)\\
.045 & 5.315(4) & .516(2) & .0047(2) & .9989(1) & .342(1)\\
.04 & 6.127(2) & .513(2) & .0048(2) & .9989(1) & .343(1)\\
.03 & 8.305(5) & .517(1) & .0045(2) & .9989(1) & .344(1)
\end{tabular}
\end{table}

\begin{table}
\caption{$N_f = 40$ Data}
\label{10.}
\begin{tabular}{dddddd}
$\beta$ & $P$ & $\langle{\overline\psi}\psi\rangle$ & $\chi$ & $m$ & $c$\\
\tableline
.05 & 3.042(4) & .1522(5) & 10.6(1) & .012(1) & .063(1)\\
.04 & 4.275(4) & .1567(4) & 11.9(1) & .024(1) & .069(1)\\
.03 & 7.648(5) & .1672(4) & 15.2(2) & .036(1) & .075(1)\\
.02 & 11.956(6) & .1686(5) & 15.9(2) & .037(1) & .089(1)\\
.01 & 23.207(7) & .1692(5) & 15.5(2) & .034(1) & .111(2)
\end{tabular}
\end{table}

\newpage

\section{Figure Captions}

\begin{itemize}
\begin{enumerate}

\item $-1/ln(m)\  {\rm vs.}\  -1/ln\langle{\overline\psi}\psi\rangle$ plot
showing critical behaviour at $\beta = .17$.

\item  Equation of State for $N_f = 8$ theory.

\item  Scaling plots, Eq. (2.3) and (2.4), of monopole percolation quantities
for the $N_f = 8$ theory.

\item  Chiral condensate $\langle{\overline\psi}\psi\rangle$, monopole
percolation order parameter $M$ and average plaquette $P$ for $N_f = 12\ {\rm
and}\ 16$ theories.

\item  Same as Fig. 4 except $N_f = 20\ {\rm and}\ 24$.

\item  Same as Fig. 4 except $N_f = 8\ {\rm and}\ 24$, but the monopole
concentration $c$ is shown as well.

\item  Monopole percolation susceptibility plots for $N_f = 8\ {\rm and}\ 24$.

\item  Same as Fig. 6 except $N_f = 32\ {\rm and}\ 40$.
\end{enumerate}
\end{itemize}
\end{document}